\documentclass[runningheads]{llncs}
\usepackage{booktabs}   
\usepackage{multirow}   
\usepackage[table]{xcolor} 
\usepackage{makecell}

\usepackage[T1]{fontenc}
\usepackage{graphicx}
\begin{document}
\title{From Mind Mapping to Dream Interpretation: Understanding Muslim Users' Engagement with GenAI Tools in Religious Context}
%

\author{Ishrat Jahan\inst{1}\orcidID{0000-0003-2474-0735} \and
Soomanib Kamruzzaman\inst{2}\orcidID{0009-0000-5185-2041} \and
K M Nazib Hossen\inst{2}\orcidID{0009-0009-3955-2041} \and
Novia Nurain\inst{1}\orcidID{0009-0007-1820-6129} \and
A. B. M. Alim Al Islam\inst{1}\orcidID{0000-0001-8159-6114} 
}
\authorrunning{I. Jahan et al.}
%
\institute{
Department of Computer Science and Engineering, Bangladesh University of Engineering and Technology\\
\email{\{ishrat,nnurain,alim\_razi\}@cse.buet.ac.bd} \and
Independent Researcher\\
\email{soomanib,nishuptonazib@gmail.com}
}
\maketitle              

\begin{abstract}
Generative Artificial Intelligence (GenAI) tools are increasingly integrated into everyday life, including sensitive domains such as religion. Prior works examined alignment with religious values and discussed the potential risks. However, to gain empirical insights into the adoption of these tools in users' religious lives, in this study, we conducted an online survey with 178 Muslim participants in Bangladesh and collected 726 prompts on religious use cases, along with open-ended responses about using these tools in religious contexts. Using reflexive thematic analysis, we identify three key roles these tools play in users’ religious lives: (1) religious assistants, (2) perceived religious scholars, and (3) faith counselors. However, participants expressed concerns about misguidance, overreliance, and violations of Islamic tradition, and reported adopting protective strategies to avoid these pitfalls. Based on these, we propose a set of design guidelines for faith-sensitive GenAI systems that would preserve religious boundaries while supporting religious practices.

\end{abstract}

\section{Introduction}
Generative Artificial Intelligence (GenAI) tools have been rapidly reshaping human-technology interaction with their powerful capabilities in reasoning, creativity, and conversation. People are integrating them into daily life across a wide range of everyday contexts, including sensitive domains such as healthcare \cite{de2025SelfCare2}, education \cite{GenAIEducation}, counseling \cite{LLMMentalHealthCounseling}, etc., as personal assistants \cite{LLMAssistant}, creative content creators \cite{capel2024studyingSelfCare}, search engines \cite{xiong2024searchEngine}, and even social companions \cite{AIsocialCompanion}. In response to this growing adoption, HCI community has engaged extensively in examining the use, benefits, and risks of these tools in various contexts. Religion is one such domain that has just recently begun to receive attention in this discussion with the increasing adoption of these tools in people’s everyday religious and spiritual practices \cite{blindFaith,Wright2024GodChatbots,Irfan2025large}. Such a trend has, therefore, made it imperative to examine how these tools are integrated into users' everyday religious practices and the implications of such usage.

Recent HCI research on religion and spirituality has emphasized the importance of considering faith as an integral part of human experience that shapes technology adoption, usage, etc., \cite{rifat_integrating_2022,forGodsSake}. In light of this, studies are beginning to consider how such faith-informed perspectives extend to the use of contemporary GenAI tools \cite{Irfan2025large,gptBible,bianaLlmMdpi,abstentionSurvey2024}. Some studies evaluated how well these tools perform or align in religious context \cite{blindFaith,bhojani2023truthAndRegret,lahmar2025islamtrust}. However, challenges with religious value alignment and users’ inclination towards accepting AI-generated religious guidance is already evident from the literature \cite{blindFaith,lahmar2025islamtrust}. Yet, it remains largely underexplored how users employ these tools in their daily religious practices, and the challenges they encounter in doing so. Gaining empirical insights into these aspects is essential, as it forms the foundation for understanding the impacts of such usage and guiding future design in this sensitive domain. To address this gap, our study aims to investigate how users adopt, perceive, and negotiate the use of GenAI tools for religious purposes. We particularly focus on Muslim users in Bangladesh in this study. We have organized our study keeping the following research questions in mind: (1) \textbf{RQ1}: How do Muslims engage with GenAI tools for religious purposes? and (2) \textbf{RQ2}: What concerns do they have, and how do they navigate them when using GenAI tools in religious contexts?

To answer proposed research questions, we conducted an online survey focusing on the ways users use, employ, and perceive GenAI tools in their religious activities. We received a total of 182 responses and retained 178 after cleanup. The survey contained questions on participants’ familiarity with GenAI, their religious adherence, experiences, and queries related to religious use, and demographic details. Participants were asked to share at least four prompts they have used or would use for religious purposes. In total, we received 726 prompts. We performed reflexive thematic analysis \cite{braun2019reflecting} on the prompts and open-ended questions to find out the underlying themes on the usage of GenAI tools for religious purposes. Our analysis reveals how Muslim users are integrating GenAI tools into their religious lives, along with their potential and pitfalls in this domain. We unpacked three distinct roles GenAI tools play in this context: as religious assistants supporting productivity and learning, as Shaikhs or Islamic scholars offering instant theological clarification and personalized religious rulings, and as faith counselors, providing emotional and spiritual support grounded in faith. However, the advantages are often overshadowed by their expressed concerns about the vulnerabilities and risks these tools might pose to one’s faith coupled with the sensitivity of this particular domain. Especially, tensions arise around the role of scholar by these tools as it conflicts with Islamic traditions and creates fear of potential misguidance. Finally, we bring forth the practical strategies participants often adopt to avoid these risks. Some of these include carefully selecting `safe-zones', i.e., low-risk Islamic domains for perceived harmless engagement, manual verification, and cross-checking responses with external trusted sources, etc., to enable safer use.

This study contributes to the intersection of religion, GenAI, and HCI literature in several ways. First, our work provides empirical evidence 
on how Muslims envision different roles of GenAI to support their religious practices. Unpacking such roles will advance the understanding of the interplay of such tools with religious practices. Second, it informs the literature about the concerns users raise regarding the use of GenAI tools in religious contexts. These concerns point to opportunities for developing safeguards, transparency mechanisms, and religiously grounded design approaches that can address these risks and facilitate responsible engagement. Third, we suggest the need for value alignment that goes beyond improving the performance of these tools in religious responses and respects boundaries, i.e., either abstain from or exercising caution when taking on specific roles or answering queries that may contradict established Islamic tradition. Lastly, building on our findings, we propose design guidelines for GenAI in religious contexts, including specialized agents for diverse use cases, scholar-informed safe and unsafe interaction zones, uncertainty indicators with source linking, and nudges to consult qualified Islamic scholars. We expect these design directions to empower users harnessing the capabilities of GenAI tools while being mindful of their limitations to promote safer, more reliable, and faith-aligned engagement.

\section{Background}
    We situate our study in the intersection of HCI, GenAI tools and religion. Prior HCI studies have emphasized on the importance of understanding how faith-based communities adopt, adapt and resist technologies \cite{rifat_integrating_2022,forGodsSake} that are shaped by their religious values, community norms, etc. Though GenAI tools are rapidly being integrated into numerous aspects of human life including religion \cite{RiseOfChatGpt}, empirical insights into how different faith-based communities engage with these tools are underexplored. 

    Despite being a sensitive and value-laden domain, GenAI tools are being employed in different religious or spiritual settings \cite{RiseOfChatGpt,kwon2024spiritualAI,cho2025shamain,gptBible}. In response, studies are delving into understanding this growing engagement with GenAI tools for religious information seeking, learning, and exploration across faith traditions \cite{bhojani2023truthAndRegret,gptBible,LLMBible}. A number of studies have also investigated their use in disseminating religious education \cite{Irfan2025large,alharbi2025llms}. Interestingly enough, a few studies  have positioned these tools as godly entity and attributed divine power \cite{Wright2024GodChatbots,ohman2024weGodLLM} to them whereas some even went further in embodying them as Gods \cite{cho2025shamain}. Collectively, this literature shows that GenAI tools are becoming deeply entangled with religious meaning-making and demands more in-depth research to understand the interplay.

    Alongside adoption, a growing body of work has critically examined value alignment \cite{blindFaith,lahmar2025islamtrust}, reliability \cite{bhojani2023truthAndRegret}, and bias \cite{llmBias,persistentAntiMuslimBias} in GenAI systems when applied to religious contexts. This is particularly crucial given the interpretive and authoritative nature of religious knowledge. Prior studies have documented hallucinations in LLM-generated religious content, including inaccuracies in Qur’anic verses and references \cite{bhojani2023truthAndRegret}. Other research shows that while Muslim users may express a preference for scholarly or authoritative religious sources, they still prefer responses from GenAI tools and often fails to identify inconsistencies or errors in AI-generated responses \cite{blindFaith}. Besides, in examining the efficacy of LLMs in Islamic context, the study in \cite{atif2025sacredAbstention} emphasized task-specific evaluation and cautious deployment of LLMs in religious contexts \cite{atif2025sacredAbstention}. On the other hand, concerns around ideological bias, cultural assumptions, and misrepresentation of religious values have also been raised \cite{persistentAntiMuslimBias,llmBias,bianaLlmMdpi}. Collectively, these studies highlight the wide spectrum of risks associated with relying on GenAI tools for sensitive religious guidance. This body of literature therefore calls for deeper investigation into how different religious communities engage with GenAI tools, in order to surface community-specific challenges, concerns, and expectations around their use in supporting faith and religious practice.
    
    In response to these limitations, recent work has increasingly focused on designing specialized religious GenAI tools \cite{mufassirLLM} and domain-specific LLMs \cite{ling2023domainSpecializationLLM} to better support religious use cases. Several such specialized tools are now publicly available, though in their early stages \cite{ansari}. Studies that surface users’ concerns, expectations, and lived experiences related to the engagement with GenAI tools for religious purposes are therefore crucial to inform the responsible evolution and refinement of these systems.

\section{Methods}
We conducted an online survey to understand Muslim users' perceptions and usage of GenAI tools for religious purposes. Our study was approved by the university's Ethical Committee. The study was conducted during the summer of 2025.

\subsection{Data Collection}
We conducted our survey using Google Forms. The survey consisted of 16 questions, including 5 open-ended questions, organized into four sections. We started our survey with a brief overview of this study. The first section asked questions related to participants' familiarity with GenAI tools. The second section focused on gathering information about respondents' religious adherence, i.e., how they identify themselves in terms of practicing Islam (e.g., fully committed, aspiring, etc.) and how long they have been practicing Islam. The third section explored participants’ perceptions and experiences with GenAI tools for religious purposes (experiences with GenAI tools, their motives for using GenAI tools for religious purposes, and their concerns regarding such use). In this section, participants were also asked to provide at least four queries they usually ask or would like to ask a GenAI tool. The last section collected demographic information, including age, gender, education, and profession. The survey data are stored on secured university servers and are accessible only to the research team.

\subsection{Participants}
To recruit participants, we adopted snowball sampling approach \cite{snowball}. We shared our study information through social media platforms and our personal networks and requested participants to share it in their own networks. A total of 182 responses were collected. We excluded 4 responses that seemed misaligned or absurd to the authors. After this data cleaning process, the final sample consisted of 178 participants. All of our participants are from Bangladesh. The gender of the participants was balanced with half male and half female, and the majority were aged between 19 and 30. The majority of participants reported being familiar with GenAI tools, with only 3.9\% indicating no familiarity. Additionally, about half (50.6\%) of the participants reported prior use of GenAI tools for religious purposes. 
The most frequently used GenAI tools were ChatGPT, followed by Gemini and others. Most participants identified as practicing Muslims, with the majority describing themselves as either fully committed or consistently practicing, and over two-thirds (67.4\%) reported practicing for over nine years. However, we did not collect information about participants’ sectarian affiliations or schools of thought. The details of participant demographics are included in Table \ref{tab:demo}.

\begin{table*}[!t]
\centering
\caption{Demographic of the participants}
\label{tab:demo}
\renewcommand{\arraystretch}{1.15}
\begin{tabular}{l l c}
\toprule
\textbf{Category} & \textbf{Description} & \textbf{Participants (N=178)} \\


\midrule
\multirow{2}{*}{Age (\#/\%)} 
& 19-30   & 119 (66.8) \\
& 31-40 & 37 (20.8) \\
& 41-50 & 16 (9.0) \\
& 51-60 & 6 (3.4) \\
& 61 or above & 0 (0.0) \\

\midrule
\multirow{2}{*}{Gender (\#/\%)} 
& Male   & 89 (50.0) \\
& Female & 85 (47.8) \\
& Prefer not to say & 4 (2.2) \\


\midrule
\multirow{6}{*}{\makecell[l]{Education Level (\#/\%)\\(highest completed)}} 
& College degree or below       & 32 (18.0) \\
& Bachelors' degree         & 89 (50.0) \\
& Masters' degree  & 49 (27.5) \\
& Doctorate degree    & 8 (4.5)  \\


\midrule
\multirow{8}{*}{\makecell[l]{Access to Technology (\#/\%)\\(multiple responses)}} 
& Laptop computer   & 159 (89.3) \\
& Desktop computer  & 121 (68.0) \\
& Smartphone        & 175 (98.3) \\
& Feature phone     & 93 (52.2)  \\
& Tablet            & 95 (53.4)  \\
& E-reader          & 25 (14.0)  \\
& Wearables         & 55 (30.9)  \\

\midrule
\multirow{8}{*}{\makecell[l]{Most frequently accessed GenAI tool (\#/\%)\\(multiple responses)}} 
& ChatGPT    & 162 (93.9) \\
& Gemini     & 100 (58.0) \\
& DeepSeek   &  86 (49.7) \\
& Copilot    &  63 (36.5) \\
& Claude     &  49 (28.2) \\
& Perplexity &  26 (14.9) \\
  
\bottomrule
\end{tabular}
\end{table*}

\subsection{Data Analysis}
We collected and compiled the sample prompts shared by participants, resulting in a total of 726 prompts. We adopted a reflexive thematic analysis \cite{braun2019reflecting} to analyze the prompts along with the qualitative responses of the open-ended questions. We read and re-read the responses to familiarize ourselves with the data. At first, the research team jointly open-coded a subset of responses and created an initial codebook through discussion. 
Next, two researchers coded the rest of the responses and modified the codebook through the process. Emerging themes were then identified by reviewing the coded data and grouping conceptually similar codes into higher-level categories. Through several synchronous discussions with the research group, these categories were consolidated into three overarching themes. 
The themes included GenAI tools taking different roles to assist participants in their religious matters, concerns around such use, and strategies for adapting to the perceived limitations of these systems.

\section{Findings}
In this section, we document the findings of our study. We delve into each theme consolidated by our analysis, with a necessary breakdown of subthemes where needed.

    \subsection{Perceived Religious Roles of GenAI tools} 
    Our analysis revealed the diverse and highly personalized ways genAI tools are being integrated into the lives of Muslims for navigating religious aspects. We categorized these uses into three subthemes based on the roles they play in the religious lives of our participants. First, these tools are mostly used as religious `assistants' as people rely on them to make religious tasks easier. Next, they use them as 'Shaikh' or religious scholar for learning about religious rulings, clarification, reasoning, etc. Finally, as a `Faith Counselor,' people seek advice, encouragement, or empathetic support from these tools. In the following sections, we elaborate each of these roles.


    \subsubsection{GenAI tool as a Religious Assistant} 
    The most common usage pattern that emerged from the prompts is the use of genAI tools as a religious assistant to avail support in daily religious worship or performing important rituals. Participants reported leveraging GenAI tools to enhance their religious productivity in multiple ways. For example, utilizing the content creation capabilities of these tools, they employed them to design structured routines for prayer, Qur’an recitation, and time management, etc., to achieve greater consistency in their daily worship. As P69 shared how she approaches for routine, ``\textit{Can you help me to create a simple daily routine to stay spiritually connected to Allah?}''

    Similarly, participants demonstrated creative ways of using genAI tools to aid their memorization and learning practices. They explored how these tools could generate visual or structural aids, such as mind maps, to scaffold cognitive processes in religious study. P8 reported,
    \begin{quote}
    ``\textit{For Qur’an memorization, I sometimes ask the chatbot to create a mind map using the meaning of the verses to help me remember the sequence.}''
    \end{quote}
    P173 even called ChatGPT her `Hifz partner' i.e., partner in Quran memorization due to the relationship built up through seeking such assistance in this context. Meanwhile, utilizing their information retrieval power, participants mentioned using these tools to search for references in the Qur’an and Hadith, or extracting historical events. Besides, some participants mentioned asking for assistance in calculating Zakat or annual obligatory donation for Muslims. 
    
    In addition, participants explored more personal forms of guidance, such as clarifying their duties at a specific situation or a specific role as a Muslims. Others even sought dream interpretation, with one P45 noting, ``\textit{[...] based on my given context, it (ChatGPT) tried to generate an Islamic explanation of the dream referring to Qur’anic quotes  [...]}''. Dawah or spreading Islam is another field mentioned by some participants, where they took help, such as brainstorming Dawah strategies with these tools.

    Beyond worship, participants also highlighted the use of GenAI tools as accessible medium for religious knowledge building and skill development such as Arabic language learning, memorization support, searching references, etc. For example, P8  explained:
    ``\textit{I usually ask small questions related to learning Arabic [...] I also use it to find the complete text of a Hadith when I only know a fragment or summary. Additionally, I use it to search for Qur’anic verses on specific topics.}''




     Seeking general religious information also emerged as a frequent use case. Many described using GenAI like a search engine to quickly extract the information they needed. Participants asked about established facts, such as the number of verses on specific topics, particular hadiths or Qur’anic verses for given situations, and historical events. P79 summarized this use as:
    \begin{quote}
    ``\textit{I think chatbots can be used for learning about Islamic history, the life of every Nabi and Rasul, especially Prophet Muhammad (SAW), how Muslim leaders used to bring justice to crimes, and general Islamic knowledge like the number of Hadith books}''
    \end{quote}
    
    Overall, participants relied on these tools not only for religious learning and practice but also for enhancing productivity and religious knowledge.




    \subsubsection{GenAI tools as Shaikh or Religious Scholar} 

    
    A prominent way people interact with GenAI tools surfaced in our findings is to seek Islamic rulings, which traditionally people asked to Shaikh or scholars. These queries ranged from generic Islamic rulings, everyday uncertainties, to serious situational queries. Some participants reported using these tools to check the validity of their worship in a given situation. Such as P46 asked ``\textit{[...] if I have lain down, but haven't completely fallen asleep, do I need to perform wudhu?}'' Similarly, P97 queried about his validity of fasting,
        ``\textit{I've accidentally eaten something although I am supposed to be fasting. Does that mean my fasting will no longer be accepted? }'' 
    
    Meanwhile, some participants used chatbots to seek guidance on the permissibility of specific actions. For instance, P117 asked, ``\textit{Is it permissible to cut tendons/legs of the sacrificing animals before they rest fully?}'' while another inquired about daily worship, such as whether it is permissible to pray the Esha prayer after 12 a.m. 

    Beyond routine practices, participants also raised more critical questions that involved exceptional circumstances where following standard rulings was difficult or not feasible. For example, one participant prompted, ``\textit{Can I keep my hair short, as a girl, not for style, fashion, or to look like a man, but because I genuinely cannot maintain it, have little time, and often fall sick from wet hair after bathing?}'' Similarly, another participant asked about the permissibility of eating pork in extreme conditions, while P88 inquired whether one could relax requirements for modest clothing in situations of illness. Some threw more serious and challenging situations to these bots and sought rulings in those contexts. P94 elaborated as below:
     \begin{quote}
        ``\textit{Can I lie and say I am not Muslim but I am in my heart... To save my family from getting murdered and getting serious harm? Or verbally saying will deem me as Kafir since verbal recognition is also part of Shahadah and faith. What should I do in this situation?}''
    \end{quote}

    Others sought clarification on everyday moral dilemmas such as, ``\textit{What does Islam say about copying or cheating in exams?}'' Similarly P73 reported sharing dilemma with these tools and asking for Islamic guidance, ``\textit{
            I am facing some trouble while choosing between two activities in my day to day life. I want a comparison between them with reference to the Quran and Islam.
        }''

    Many participants mentioned seeking logical reasoning or clarification of any ruling or event from GenAI tools. For example, P104 prompted as, ``\textit{If Allah knows about the future, how can a human change the future that is known by Allah?}'' While P49 posed something more profound, ``\textit{What makes Qur’anic verses resonate emotionally and intellectually with both unlettered individuals and highly educated scholars?}''. Many participants also reported using GenAI tools to explore answers to controversial questions that are frequently raised either within Muslim communities or by non-Muslims. These included sensitive topics such as, ``\textit{why women are oppressed in Islam}'' or broader claims like ``\textit{why Islam is the solution for all''}. For these users, GenAI tools served as a convenient space to articulate and test responses to criticisms of their faith, or to strengthen their ability to explain Islamic positions in interfaith or public discussions. 
    
     Lastly, some participants shared the motivations behind relying on these tools for religious purposes. P112 mentioned that the perceived reliability and trustworthiness of these tools rooted from their vast knowledge base that hold access to far more religious content than an ordinary individual could. She explained:
     
     \begin{quote}
    ``\textit{As I am not an Islamic scholar, I do not know most of the Hadiths and interpretations of Quranic verses. As LLMs are fed with almost all the data available on the internet, these systems read all the Hadiths and Verses. So, I usually use them to find answers to some of such questions using chatbots. 
    }''
    \end{quote}

    Many others mentioned ease of use and accessibility behind such usage. On the other hand, some even asked GenAI tools to emulate the persona of a religious scholar to make the interaction feel more realistic.

    \subsubsection{GenAI tool as a Faith Counselor} 
    Beyond seeking assistance in religious practices, many participants mentioned turning to GenAI tools for spiritual counseling. For these participants, these tools functioned beyond an information provider and as an empathetic conversational partner that could offer encouragement and emotional support grounded in their faith. In this regard, P74 shared an instance of how he approached these tools in similar cases:
     \begin{quote}
        ``\textit{I am going through a period of extreme self-doubt and lack of confidence because of not being able to perform as good as others in my job/academics. Can you find me some words of Allah and His prophet (Pbuh) that are relevant to my situation and can help lift me up?}''
    \end{quote}
    
    However, the attraction to this form of companionship lay in the these tools' ability to provide empathetic support without guilt-shaming. P165 explained this as, ``\textit{(These tools are) non-judgmental yet firm on principle talking space. Doesn't encourage haram, but doesn’t guilt shame either.}''
    
    Some even mentioned consulting chatbots on important life events such as marriage or handling relationships, parenting, etc., to get religious guidance on these. For example, one of the prompts of P45 was ``\textit{Help me to take mental preparation of marriage based on Islam as I'm always nervous of hearing it.}'' Some questions often reflected real-life struggles where participants wanted to align their responses with Islamic principles. For example, P35 prompted for guidance on handling interpersonal challenges in a religious context.
    \begin{quote}
        ``\textit{ How to gracefully deal with a problematic person who is trying to put me in an embarrassing situation, targeting me only because of my faith (with details of exactly what happened)?}''
    \end{quote}

    Another form of support participants sought from these tools was self-enhancement through reflection and personal growth. They described using GenAI tools to strengthen their spiritual practices, improve emotional regulation, and cultivate discipline in daily life. For example, some participants asked for tips on how to maintain focus during Salah, while others sought guidance on controlling anger, lust, or other habits they wished to abandon. GenAI tools were also consulted for relevant Qur’anic verses related to building particular character traits, as well as strategies to develop self-discipline and maintain consistency in worship. P91 elaborated this motivation in his response:
    \begin{quote}``\textit{
        I would use chatbots to support my religious activities by getting inspiration and guidance for personal growth and daily spiritual habits. For example, I might ask for short Qur’anic verses or hadiths to reflect on each day, or request stories that encourage qualities like patience, gratitude, or forgiveness. A chatbot could also help suggest ways to stay connected with my faith [...] Additionally, I’d be interested in learning how to build stronger, more compassionate relationships within my community using faith-based values.
    }''\end{quote}

    Thus, participants used these tools to stay spiritually motivated and get comfort at turbulent periods through Islamic guidance. These tools provided convenient and consistent companionship and support for aligning life challenges with Islamic principles.

    \subsection{Concerns around Interaction for Religious Purposes} 
    Our participants mentioned a number of concerns related to interacting or consulting with GenAI tools for religious purposes. Primarily, these concerns are rooted from the lack of trust or reliability of GenAI tools' responses in religious matters, as well as due to the frustration with unsatisfactory answers, and fears that inaccurate guidance could potentially harm their faith. In the following sections, we document the concerns of our participants in this context.


    \subsubsection{Being Misguided}
    The strongest concern that surfaced in our study regarding the religious use of GenAI tools was the fear of being misguided in religious matters. This fear amplified by many participants’ observations that these tools hallucinate frequently i.e., present inaccurate, misleading, incomplete, or even fabricated information about sacred texts and religious rulings. In this regard, P78 indicated that such inaccuracies could create confusion, while P69 expressed a deeper concern that these errors may lead to deviation from Islam and ultimately cause individuals to drift away from their faith. Several participants mentioned additional issues such as \textit{misleading reference numbers, shortening main verses and hadith} as mentioned by many participants that made them deeply uneasy. While some of the participants had firsthand experience of such incident, some others did not, though are afraid of encountering so. P73 expressed her fear of getting fabrication verses of Quran as  ``\textit{
          I am afraid they may fabricate some ayaat from the Quran or they might provide some false Hadiths while providing references. 
        }''

    Participants emphasized much on the fact that religious guidance must come from sources grounded in authentic scholarship. For them, the danger lies in GenAI tools' offering generalized or misinterpreted answers. As P28 commented that ``\textit{Sometimes they share self-generated wrong information confidently}'', which could ultimately misguide believers if taken at face value. P69 urged on the alignment of guidance with trusted religious sources as below:

     \begin{quote}
     ``\textit{
           When it comes to faith, it's important that the guidance aligns with trusted religious sources and is not misinterpreted or generalized. If a chatbot provides an answer that’s not based on sound scholarship, it could be misleading.
    }''
    \end{quote}

    According to others, misguidance can take place as GenAI tools miss religious perspectives. Since they do not understand the nuances of religious questions and often lack deep religious understanding, they generate answers that are not acceptable. P135 shared such an instance as stated, ``\textit{
            [...] During this Ramadan, I saw a post in a girls’ group where it was mentioned that ChatGPT advised drinking water before iftar to keep calm. 
        }''
    In Islamic tradition, during Ramadan one must fast until the time of Iftar (the meal taken after sunset). Drinking water before Iftar would directly invalidate the fast. Such mishandling of religious rulings highlights the potential danger of relying on GenAI tools for sensitive matters of faith. For this reason, Many of the participants have expressed their deep concerns about these tools providing Islamic rulings.  Participant P68 has commented in this regard:
        \begin{quote}``\textit{
            [...] can such agent actually understand the subtleties of religious scriptures when even most people can't understand them properly and misinterpret them according to their need? 
        }''\end{quote}  

    Participants also expressed concern about GenAI tools' regarding the understanding of the complexity of Islamic jurisprudence and the possibility of disregarding essential considerations (such as school of thought, difference of opinions, etc.). Many expressed  fear that reliance on these systems for rulings could mislead ordinary Muslims, especially those with little prior knowledge to verify the information they receive. One participant noted, ``\textit{It is unclear whether chatbots take into account considerations such as madhhab, ijma, or qiyas. Without being sure of this, I believe that relying on chatbots for religious rulings, knowledge, or guidance is highly risky.}''

    Other than the inaccuracies in the information itself, many participants have reported inconsistency in the responses as a concern. Inconsistency may include - sycophancy, manipulation and misalignment of context due to aggregation from multiple sources.  Some, other participants have expressed concerns that bots tend to respond in a way that pleases the user as P63 reported  that  ``\textit{Often, the responses seem to be influenced by what the user wants to hear, rather than providing a truth-based answer. }''


    Thus the risk with sycophancy, i.e., GenAI tools aligning their answers to satisfy users in religious context can lead to some serious consequences as mentioned by our participants. P157 highlighted how this can be misused by many ``\textit{AI just don't want to break the user’s heart, so people may use AI to act according to their own wishes and then believe that what they did is correct.}'' This eventually risks deviating people away from the right track as expressed by our participants.



    \subsubsection{Deviation from Traditional Learning}
    A prominent theme that emerged from participants’ responses was their concern that over-reliance on GenAI tools for religious purposes could gradually drift people away from traditional modes of learning Islam. Several participants highlighted the fear that people might increasingly turn to ``Shaikh ChatGPT'' or ``Shaikha Gemini'' for guidance on sensitive issues like Fiqh or Islmaic jurisprudence. The traditional mode as some of them mentioned to be to consult `people of knowledge' i.e., scholars, knowledgeable individuals, etc., citing the Qur’anic command to consult ``people of knowledge,'' participants emphasized that people should go to them for any queries regarding religious topics. Many other participants urged that to know religion, people should earn knowledge on Quran, Hadith, read religious books and seek answers from scholars. Therefore, according to them, bypassing scholars in favor of GenAI tools undermines the Sunnah practice of seeking guidance from living scholars. P89 commented on this as  ``\textit{
          Religious practices are guided by Sunnah (The practices of Prophet Muhammad (Pbuh)). Asking question to a living person is Sunnah, asking to a machine (Chatbot) is not Sunnah. 
    }''

    
    Another recurring sentiment was that while GenAI tools might be helpful for retrieving information, it cannot replace the Hikmah i.e., wisdom of human scholars. P101 commented in this regard that ``\textit{These are machine learning models. They will act based on how they are trained. Machine will never think like a human being!}''. Participants also worried about the theological implications of such reliance. P171 pointed to the danger of attributing undue authority to GenAI tools as ``\textit{people are taking AI tools as a God these days}''. 

    Finally, several participants also expressed concern that younger generations may normalize relying on GenAI for religious guidance and stop verification and scholarly consultation. In this regard, P69 expressed her fear:

    \begin{quote}``\textit{
          Lastly, there's a concern that people especially the younger generation might over-rely on chatbots and stop turning to real scholars or their local religious communities for guidance, which can weaken communal bonds and the traditional ways we learn and grow in our faith. 
    }''\end{quote}

    \subsubsection{Bias and Exploitation}
    





    Participants voiced deep concerns about the inherent bias within GenAI tools that may disrupt their religious engagement. In religious contexts, some participants mentioned these tools having ``\textit{too much liberal views}''. Many explicitly mentioned their discomfort of using these systems as they are ``\textit{made by Non-Muslims}''. They viewed these technologies as fundamentally compromised by the worldviews and agendas of their creators and operators and would prevent them from disseminating religious information neutrally. P7 commented in this regard:
    \begin{quote}
        ``\textit{Most of the chatbots are biased on the ideologies of the country or group of developers who create and operate them. So, they answer according to their own way of bias which can provide misleading answers. }''
    \end{quote}

    P68 even mentioned that the bias would remain even if a specialized version is made for religious purpose if not explicitly built from scratch. He commented as, ``\textit{Yes, if they are a modified and specialized version of existing chat bot (not explicitly built from scratch), they will definitely have bias introduced by the foundational model. }'' Another dimension of concern expressed by participants was the fear of exploitation through GenAI tools. Some participants, suspected that these systems could be weaponized by some powerful actors of the world, to gradually distort or weaken the faith of Muslims. P104 shared her concern as :

    \begin{quote}
        ``\textit{Some of those bots have been made by Zionists so they may try to pollute out faith covertly.}''
    \end{quote}

   Some participants also worried that by observing the types of struggles Muslims share with GenAI tools, these systems could learn sensitive patterns that may be misused against Muslim users. P35 apprehended that hostile powers could build a database of vulnerabilities and use it to manipulate, desensitize, or distract the ummah. One participant went as far as to suggest that ``\textit{If Dajjal (the false Messiah) were to use this technology, he would get a good database on people’s weaknesses,}''. This reflects deep unease about the potential for such tools to be exploited in ways that harm Muslim users.

    \subsection{Strategies for Safe Engagement} 
    To harness the benefits of GenAI tools while remaining mindful of their potential risks, participants adopted a range of strategies to avoid or mitigate risky interactions. We detail these strategies in the sections that follow.







    \subsubsection{Careful Domain Selection}
    The most prominent strategy adopted by participants was careful domain selection, wherein they explicitly distinguished between topics appropriate for consultation with GenAI tools (safe zones) and those they deliberately avoided (danger zones). Interestingly, the definition of these perceived `safe zones' varied across participants. A majority of participants identified Fiqh, including Islamic rulings and jurisprudential matters, as a primary danger zone and consciously refrained from engaging GenAI tools in these areas. According to them, such topics required scholarly expertise that participants felt that GenAI tools do not possess. Instead, participants reported leveraging GenAI tools in domains they considered comparatively low risk, such as Arabic language learning or locating Quranic verses through keyword-based searches. As participant P48 reported:

    \begin{quote}
         ``\textit{I do not search for Islamic ruling (I mean fatwa, which may differ among the schools of thought). Rather I use them for mostly learning Arabic vocabulary, Arabic grammar, and finding verses of the Quran using keywords.}''
    \end{quote}
    
    Some participants felt that asking these tools about topics they had no prior knowledge of could be risky out of fear of misinformation. Therefore, the areas where they thought to have enough knowledge, seemed safe for them. P53 shared her thoughts 
    ``\textit{I would use it only for queries I already know well, because religious belief is sensitive. Since chatbots are probabilistic, they can sometimes be wrong, and the answer may conflict with my beliefs. It may also be that the chatbot is correct, and I lack knowledge about that topic.}''
    
    On the other and, some participants felt that consulting GenAI tools for historical information was acceptable, but cautioned against seeking guidance on matters requiring deep religious knowledge. P100 commented on this:
    ``\textit{I think if we want to know historical events from chatbot, that should be okay. But it should not comment on something that needs pure understanding of religious knowledge, as it might create huge controversies.}''

     Apart from these, some others noted that GenAI tools could be helpful to know any information where there is ``no controversy'' or simple ``fact-based information'' such as quick references to the Quran or Hadith, avoiding any religious verdict from these chatbots. In the same spirit, P72 mentioned using chatbots only for organizing time, routines, or habits while avoiding asking for any Islamic rulings. P110 recalled occasional usage of chatbots to learn about other religions. He articulated as, ``\textit{I have very rarely used chatbots in this context. [...] But I have sometimes used chatbots to look up topics in other religion or mythology unknown to me due to my interest in history.}''
    

    Overall, participants’ selection of safe domains stem either from the belief that GenAI tools would be sufficiently reliable in those areas, or would not cause spiritual harm, or from a preference for topics with little controversy.

     \subsubsection{Verification and Cross-Checking}
    Apart from domain selection, some participants mentioned adopting some additional precautionary steps. Some reported using GenAI tools for religious information search maintaining lowered expectations regarding the accuracy and authenticity of the generated responses. As P23 commented ``\textit{[...] It’s not a religious person, it’s a bot. so we should keep low expectation about accuracy of Islamic information.}'' Some participants reported that they used verification strategies to ensure reliability of the information provided by the GenAI tools. They either double-checked the information against their trusted sources or requested comprehensive references to the tools instead of accepting their responses as final authorities. This approach positioned GenAI tools as preliminary research aids or starting point of exploration rather than definitive guidance. P62 shared his strategy of cross-checking as: 
    

    \begin{quote}
         ``\textit{(I) Ask regarding a specific ayat or hadith if it exists, as in social media people may post anything tagging it as an ayat or hadith. Then I cross check with the actual book as I also don't trust chatbots for religious issues.}''
    \end{quote}

    Besides, P63 mentioned verifying the authenticity of the information by cross-referencing reliable web sources and ensuring that the content aligns with credible references. Similarly, P117 takes only those information that chatbot can provide with ample references. He said:
    \begin{quote}
         ``\textit{(I take) Any knowledge with concrete references, with sufficient options for cross-checking with all sources (including source tree) as I find that authenticity is one of the most critical missing pieces in Islam these days.}''
    \end{quote}

    A few participants mentioned that, when asking for rulings, they provided additional information such as their preferred scholars, school of thought, etc., to provide enough context to these tools and ensure the responses remain aligned with their way of belief. P146 shared her style of prompting as 
         ``\textit{I often ask for fatwas because they provide quick and well-organized answers. However, I follow the scholars of SeekersGuidance and I am a follower of the Hanafi madhhab, so I mention both of these when asking for a fatwa. }''
    
    On the other hand, some mentioned taking help from genAI tools only when they have no reliable alternatives. Such as P48 shared her way of taking help as, ``\textit{For Arabic grammar, it is not reliable at all. 
    So personally I seek AI's help in this regard only when my teacher is not available.}''

    To summarize, these strategies present a sophisticated interaction approach with GenAI tools to balance the benefits while remaining aware of the risks associated with their adoption.

\section{Discussion}
This study provides empirical depth to the discussion of GenAI use in religious contexts. In this section, we discuss the implications of our findings and propose design guidelines. 



Our findings reveal a number of important aspects regarding usage of GenAI tools in religious context. The three distinct roles, as identified from our participants' use of these tools in their religious lives, imply that they aim to leverage their extended capabilities to mediate religious engagement in ways that are more personalized, accessible, and responsive than conventional sources. Faster access to information in natural language, along with content creation and brainstorming support, enables participants to ease and extend their religious learning and practice and use them as `assistants'. Moreover, requesting religious rulings, and seeking clarifications on complex theological concepts are made possible in real-time through these tools. 
Besides, seeking spiritual support during difficult times through Islamic perspectives, receiving encouragement to strengthen religious commitment, etc., positioned these tools in the role of a faith counselor. 
These roles demonstrate that the use of GenAI tools in religion is situated rather than uniform, as users dynamically shift between different modes of engagement depending on their immediate needs. Some earlier studies reported similar findings that also uncovered diverse roles of these genAI tools in other domains \cite{capel2024studyingSelfCare}. Therefore, a rigid, one-size-fits-all persona is inadequate for the lived realities of faith. We, therefore, suggest designing specialized agents aligning with these identified roles that can be invoked based on the user's dynamic needs. These agents would not only adapt to user-specific contexts (e.g., sectarian affiliations or contemporary situations) but also provide proactive support, such as consistency checking, practice reminders, etc. These are the tasks where general-purpose tools often falter. Some earlier studies suggest changing these tools' behaviour as a chameleon and adopt to the specific usage context \cite{desai2024examining}. Similarly, we suggest that in the context of religion, a more flexible approach to design, where multiple metaphors or roles can co-exist or be switched dynamically, could better reflect the lived realities of religious engagement.

However, the concerns raised by participants underscore that hallucinations, misinformation, and bias in GenAI can be especially consequential in religious contexts. Especially, participants expressed grave concern about these tools taking the role of scholars and causing overreliance on them for religious rulings and advice. They feared that these tools lack of depth of training in Islamic jurisprudence and the nuanced reasoning and, limited awareness of sectarian and other distinctions, inherent Western bias, possibility of data exploitation, etc. These fears are relevant and have been documented in prior studies across other sensitive contexts in the literature \cite{de2025SelfCare2,capel2024studyingSelfCare}, including religion too \cite{blindFaith,bhojani2023truthAndRegret,Wright2024GodChatbots}. Besides, the representational biases embedded in LLMs \cite{kotek2023genderbias,llmBias} risk not only marginalization of Islam but also within Islam by suppressing the voices of certain schools of thought or diminishing legitimate differences in opinion due to a lack of representational data \cite{llmBias}. Besides, fear of targeted disinformation or exploitation by adversarial entities through these tools are also evident in the literature, particularly in the sensitive domains such as health where genAI can enable the inexpensive creation and dissemination of targeted disinformation at scale \cite{safety_ai_health}. Interestingly, our study revealed that even if these tools were always accurate and safe, concerns still remain about deviation from the established Islamic tradition of seeking guidance from qualified scholars. We also documented the protective strategies our participants adopt to navigate these risks and leverage the extended capabilities. These findings provide avenues for designers to improve the experience of users in this context through design.

To address these risks, we propose several design pathways building on our findings while drawing inspiration from the literature. While we strongly support the call for value alignment and domain specialization of GenAI tools in religious context, we argue that these alone can neither guarantee error-free outputs nor fully address deeper concerns rooted in Islamic tradition. In fact, these tools are unlikely to become entirely free of hallucinations in the near future as predicted by earlier studies \cite{leiser2024hill}. Even if these tools are completely reliable and safe, delegating religious authority to them and overreliance on these tools go against Islamic tradition. Building on these concerns, we provide some alternative directions that designers can tap into for the design of these tools, respecting boundaries of faith. First, formalizing safe and unsafe zones of interaction in this domain can be helpful for users to interact with confidence and avoid risky interactions. Here, domain experts, i.e., religious scholars, can help draw boundaries between permissible and risky or prohibited interactions with these tools so that users remain informed about acceptable use as their own perception of what is safe may not always match the actual risks as suggested by earlier studies \cite{blindFaith}. Prior works also emphasize the value of expert-curated guardrails to guide what systems can or cannot generate in high-stakes contexts \cite{expertGuardrails}. Therefore, systems can enforce these boundaries through design. For queries that fall within unsafe or impermissible zones, GenAI systems should deliberately abstain from answering as done in the literature \cite{abstentionSurvey2024,contextualNoncompliance2024} and nudging users to consult qualified scholars instead. In doing so, they can explicitly acknowledge the religious significance of adopting such action. However, complete abstention may not be appropriate in all contexts, such as for reverts or users with limited access to scholarly guidance, where GenAI can offer meaningful support \cite{Irfan2025large} that has been echoed in our findings too. Therefore, when responding to sensitive or high-stakes religious queries, GenAI tools should adopt conservative response strategies, such as providing high-level guidance, clearly communicating uncertainty, issuing explicit disclaimers, or providing sources and references to enable verification and cross-checking. These system-level supports to detect and communicate uncertainty can help users navigate potential hallucinations as established in the literature \cite{leiser2024hill,nahar2024fakesOfVaryingShades}. Besides, they can present multiple scholarly perspectives without asserting authority or can ask the user for further information, such as the school of thought or preferred scholars, to better understand the context for improved response. Overall, our study emphasizes that while generative AI holds promise in the religious context for general people, its deployment must be carefully bounded and grounded by religious values.

\section{Conclusion and Future Works}
To conclude, this study unpacks the multifaceted ways Muslim users utilize GenAI tools into their religious lives. We document enthusiastic adoption of these tools in various religious practices as religious assistants, scholars or faith counselors as well as cautious or restrictive usage for safeguarding against overreliance and protecting the integrity of their faith from the well-documented limitations and vulnerabilities of these technologies. These insights are expected to enrich HCI research on religion and AI by bringing forth the lived experiences of Muslim users with GenAI tools. Moreover, they highlight the need for designs that preserve faith boundaries while allowing them to leverage the extended capabilities of these tools and experience greater ease and support in their religious practices. The design guidelines provided in this study would illuminate the paths of future researchers and designers in designing GenAI tools that ensures safer, more reliable, and faith-aligned
engagement. 

Looking ahead, we plan to extend our research in several ways. First, as our sample is limited to Muslims in Bangladesh, future studies that examine Muslim communities across diverse cultural contexts and sectarian or denominational traditions may provide a richer understanding of how theological and cultural differences influence the use of GenAI tools in religious practice. Next, longitudinal studies can be conducted to track how usage patterns, concerns evolve over time and uncover factors that impact the usage. On the other hand, collaboration with Islamic scholars to define permissible interaction scopes would be a promising direction for providing users with more confidence in their interactions. Furthermore, participatory approaches that involve multiple stakeholders, such as users, scholars, and designers, would be valuable to translate religious values and concerns into actionable design requirements and interaction guidelines aligning with the lived religious needs of Muslim communities.



\bibliographystyle{splncs04}
\bibliography{sample-base}
\end{document}